# On size and growth of cells.

Arezki Boudaoud

*Department of Mathematics, MIT, 2-335, Cambridge MA02139, USA and Laboratoire de Physique Statistique de l'ENS, 24, rue Lhomond, 75231 PARIS, France*

**Understanding how growth induces form is a longstanding biological question[1-6]. Many studies concentrated on the shapes of plant cells[2], fungi[3] or bacteria[3-5]. Some others have shown the importance of the mechanical properties of bacterial walls[5] and plant tissues[6] in pattern formation. Here I sketch a simple physical picture of cell growth. The study is focussed on isolated cells that have walls. They are modeled as thin elastic shells containing a liquid, which pressure drives the growth as generally admitted for bacteria[4] or plant cells[7]. Requiring mechanical equilibrium leads to estimations of typical cell sizes, in quantitative agreement with compiled data[8-27] including bacteria, cochlear outer hair, fungi, yeast, root hair and giant alga cells.**

The starting point is a crude physical description of a cell (fig. 1**a**). A liquid (the cytoplasm) is contained in a thin elastic shell (the cell wall). The physical parameters involved are the cell radius of curvature $R$, the wall thickness $h$, the elastic modulus of the wall material $E$ and the pressure $P$ exerted on the wall (or the turgor pressure). $P$ is the difference between pressures inside and outside the cell. In this Letter, I derive scaling laws relating these parameters and test them with experimental data.

The analysis relies on the following remarks. The turgor pressure and the thickness of the wall are mainly regulated by the cell physiology. Wall growth is similar to plastic deformations as established for plant cells[7]: The wall behaves as an elastic material below a critical strain $a_y$ and grows above by yielding to stress. So, the wall is modeled as a perfectly plastic material[28], which yields in extension and not in compression (see fig. 1**b**). The cell can also regulate the wall plasticity[7]. I assume that the wall has no intrisic (natural) curvature when it is formed. Finally, the growth is slow: The characteristic time for growth is much larger than the time needed to reach mechanical equilibrium. Consequently, the cell is considered to be at mechanical equilibrium.

I first estimate the cell mechanical energy, which is minimal at equilibrium. A thin shell has two modes of deformation: Stretching and bending[29]. The stretching energy is proportional to the strain squared, but if the material undergoes plastic deformations (see fig. 1**b**), then energy is stored only below the yield strain $a_y$, so the stretching energy (per unit surface) scales as $E_s \sim E h a_y^2$. The elastic energy (per unit surface) for bending is proportional to the square of the mean curvature, $E_b \sim E h^3/R^2$. In bending, the outside half of the wall (with respect to the center of curvature) is elongated while the inside half is compressed. When plastic flow occurs, it is restricted to the outside of the wall, because the material is considered to yield only in extension (fig 1**b**). In this case the effective thickness is reduced in the bending energy by factor of about 2: Its order of magnitude remains the same. The potential energy (per unit surface) corresponding to the turgor pressure is proportional to the volume/area ratio, $E_t \sim PR$.

The yield strain for most materials is smaller than of $10^{-2}$, while $h/R$ for most cells varies between $10^{-2}$ and $10^{-1}$. So, it is reasonable to consider the limiting case when $a_y \ll h/R$, so that $E_s \ll E_b$. In this case, a characteristic size for cell sizes results from the balance between bending and pressure ($E_b \sim E_t$),

$$R = \alpha h (E/P)^{1/3}. \quad (1)$$

A simple calculation assuming that the wall adopts locally the shape of a sphere of radius $R$ and using the equation for the bending of thin plates[29] gives $\alpha \sim 1.3$ (the exact value depends on the Poisson ratio of the material). The best fit to the compiled experimental data (fig. 2**b**) gives $\alpha = 4.0$, of the same order of magnitude. The agreement is good, although I have neglected the anisotropy of the wall, which is probably important in pattern generation.

One might notice a departure from the law (1) at small radii in fig. 2. This motivates the study of the opposite limit $a_y \gg h/R$ for bacteria and cochlear hair cells. Indeed, experiments by Koch (see ref. 5) have shown that the typical strain for *bacillus subtilis* is $a_y = 0.45$. In this limit, stretching balances the turgor pressure. The longitudinal tension for a cylindrical shell[5, 29] is $T = PR/2$. If the wall yields then $T = Eha_y$ (using fact that the Poisson ratio is small for a strongly anisotropic polymer as peptidoglycan[5]). So the cell typical radius is

$$R = \beta h (E/P). \quad (2)$$

Here $\beta = 2 a_y \sim 0.9$, in agreement with $\beta = 1.0$ obtained from the fit to the experimental data (fig. 2**b**). Here I have implicitly assumed that the yield strain $a_y$ varies very little for this class of cells. I have neglected the possible electric charges, which could be important in the wall shape[5]. The elastic tension $Eha_y$ is probably the origin of the tension used in the Surface Stress Theory[4] to describe the shape of bacteria.

To summarize, a preferred radius of curvature for growing cells has been determined from the balance between the turgor and elastic forces (bending or stretching depending on the wall yielding properties). A possible physical picture is that isolated cells grow more or less spherically until this preferred radius is reached. Then they adopt a cylindrical shape to keep this radius. This remark could account for the widespread existence of isolated elongated cells: root hairs, pollen tubes, fungal hyphae, alga cells… One can also notice that the scaling laws (eqs. 1-2) roughly correspond to tip growing cells and cells undergoing diffuse growth respectively. The present results could motivate (and would need) new measurements of the elastic properties of cell walls. They would be the basis for a quantitative theory for cell growth.

**Methods**

The wall elastic modulus $E$, the turgor $P$, the cell radius $R$ (the radius of the cylindrical part for elongated cells) and the wall thickness $h$ were collected as described in the following. The two lengths were either measured from the published photographs or taken from the references. When some data were missing in the literature, they were replaced by data from similar species as indicated below. *Chara corallina*[8], except $P$[9]. *Nitella opaca* and *flexilis*: $E$[12], $P$[10, 11], $h$ and $R$[10, 12]. *Acetabularia acetabulum*: $E$=500 MPa (same order of magnitude as in refs. 8 and 10), $P$=0.5 MPa (same order of magnitude as in refs. 9 and 11), $h$ and $R$[13]. *Arabidopsis thaliana* root hairs: $E$=500 MPa[8, 10], $P$[14], $h$ and $R$[15]. *Saccharomyces cerevisiae*[16]. *Phycomyces* (sporangiophores): $E$[17], $P$[18], $h$ and $R$[19]. *Saprolegnia ferax*: $E$=2000 MPa (same order of magnitude as in ref. 17), $P$[20], $h$ and $R$[21]. Cochlear outer hair cell[22] (guinea pig). *Magnetospirillum gryphiswaldense*[23]. *Bacillus subtilis*: $E$[5], $P$[24], $h$[5] and $R$[4]. *Escherichia coli* and *Pseudomonas aeruginosa*[25]. *Saccharopolyspora erythraea*[26], except $P$=3.5 MPa (order of magnitude given in ref. 23). *Microcystis spirillum* and *Anabaena flos-*

*aquae* gas vesicles[27] (the gas pressure inside the vesicle $P=27$ kPa was estimated from the aging of critical pressure distributions in fig. 2 of ref. 27).

1. Thompson, D. A. On Growth and Form (Cambridge University Press, Cambridge, 1942).

2. Green, P. B., Cell morphogenesis. *Ann. Rev. Plant Physiol.* **20,** 365–394 (1969).

3. Koch, A. L., The problem of hyphal growth in streptomycetes and fungi. *J. Theor. Biol.* **171,** 137–150 (1994).

4. Koch, A. L., The surface stress theory of microbial morphogenesis. *Adv. Microb. Physiol.* **24,** 301–366 (1983).

5. Thwaites, J. J. & Mendelson, N. H., Mechanical behavior of bacterial cell walls. *Adv. Microb. Physiol.* **32,** 123–125 (1999).

6. Steele, C. R., Shell stability related to pattern formation in plants. *J. Appl. Mech.* **67,** 237–247 (2000).

7. Cosgrove, D., Biophysical control of plant cell growth. *Ann. Rev. Plant Physiol.* **37,** 377–405 (1986).

8. Toole, A. T., Gunning, P. A., Parker, M. L., Smith, A. C. & Waldron K. W. Fracture mechanics of the cell wall of *Chara corallina*. *Planta* **212,** 606-611 (2001).

9. Proseus, T. E., Ortega, J. K. E. & Boyer J. S. Separating growth from elastic deformation during cell enlargement. *Plant Physiol.* **119,** 775–784 (1999).

10. Kamiya, N., Tawawa, M. & Takata T. The relation of turgor pressure to cell volume in *Nitella* with special reference to mechanical properties of the cell wall. *Protoplasma* **57,** 501-521 (1963).

11. Green, P. B., Erickson, R. O. & Buggy, J. Metabolic and physical control of cell elongation rate: *in vivo* studies in *Nitella. Plant Physiol.* **47,** 423–430 (1971).

12. Probine, M. C. & Preston, R. D. Cell growth and the structure and mechanical properties of the wall in internodal cells of *Nitella opaca*. II. Mechanical properties of the walls. *J. Exp. Bot.* **13,** 111–127 (1962).

13. Dumais, J. & Harrison, L. G. Whorl morphogenesis in the dasycladalean algae: the pattern formation viewpoint. *Phil. Trans. R. Soc. Lond. B* **355,** 281–305 (2000).

14. Lew, R. R. Pressure regulation of the electrical properties of growing *Arabidopsis thaliana* root hairs. *Plant Physiol.* **112,** 1089–1100 (1996).

15. Galway, M. E., Lane, D. C. & Schiefelbein, J. W. Defective control of growth rate and cell diameter in tip-growing root hairs of the *rhd4* mutant of *Arabidopsis thaliana. Can. J. Bot.* **77,** 494–507 (1999).

16. Smith, A. E., Moxham, K. E. & Middelberg, A. P. J. Wall material properties of yeast cells. Part II. Analysis. *Chem. Eng. Science* **55,** 2043–2053 (2000).

17. Ahlquist, C. N. & Gamow, R. I. *Phycomyces*: mechanical behavior of stage II and stage IV. *Plant Physiol.* **51,** 586–587 (1973).

18. Ortega, J. K. E., Zehr, E. G. & Keanini, R. G. In vivo creep and stress relaxation experiments to determine the wall extensibility and yield threshold for the sporangiophores of *Phycomyces. Biophys. J.* **56,** 465–475 (1989).

19. Ahlquist, C. N., Iverson, S. C. & Jasman, W. E. Cell wall structure and mechanical properties of *Phycomyces. J. Biomechanics* **8,** 357–362 (1975).

20. Money, N. P. & Harold, F. M. Two water molds can grow without measurable turgor pressure. *Planta* **190,** 426–430 (1993).

21. Heath, I. B. & Kaminskyj, S. G. W. The organization of tip-growth-related organelles and microtubules revealed by quantitative analysis of freeze-substituted oomycete hyphae. *J. Cell Sci.* **93,** 41–52 (1989).

22. Brownell, W. E., Spector, A. A., Raphael, R. M. & Popel A. S. Micro- and nanomechanics of the cochlear outer hair cell. *Annu. Rev. Biomed. Eng.* **3,** 169–194 (2001).

23. Arnoldi, M. *et al.* Bacterial turgor pressure can be measured by atomic force microscopy. *Phys. Rev. E* **62,** 1034–1044 (2000).

24. Whatmore, A. M. & Reed, R. H. Determination of turgor pressure in *Bacillus subtilis:* a possible role for $K^+$ in turgor regulation. *J. Gen. Microbiol.* **136,** 2521–2526 (1990).

25. Yao, X., Jericho, M., Pink, D. & Beveridge, T. Thickness and elasticity of gram-negative murein sacculi measured by atomic force microscope. *J. Bacteriol.* **181,** 6865–6875 (1999).

26. Stocks, S. M. & Thomas, C. R. Strength of mid-logarithmic and stationary phase *Saccharopolyspora erythraea* hyphae during a batch fermentation in defined nitrate-limited medium. *Biotech. Bioeng.* **73,** 370–378 (2001).

27. Walsby, A. E. The mechanical properties of the *Microcystis* gas vesicle. *J. Gen. Microbiol.* **137,** 2401–2408 (1991).

28. Johnson, W. & Mellor, P. B. Engineering Plasticity (Ellis Horwood, Chichester, 1983).

29. Love, A. E. H. A Treatise on the Mathematical Theory of Elasticity (Dover, New York, 1944).

I am indebted to G. Toole and A. Smith for details on their data. I am grateful to Y. Couder and M. Ben Amar for getting me interested in plant and tumor growth.

**Correspondence and requests for materials should be addressed to A.B. (e-mail: boudaoud@lps.ens.fr).**

Figure 1 Theoretical setting for cell growth. **a,** Scheme of a model cell (typical radius $R$ and wall thickness $h$). Growth is driven by the inner pressure $P$. **b,** Stress-strain ($\sigma$, continuous line) and growth rate ($G$, dashed line) curves for the cell wall, which is assumed to be a perfectly plastic material yielding only in extension. The wall is elastic (with modulus $E$) for a strain smaller then the yield threshold $a_y$ and it grows above. If the stress is decreased, the released elastic energy (the shaded area) is $1/2 \cdot E a_y^2$.

Figure 2 Experimental testing of the scaling laws for cell radii. **a,** Cell radius $R$ as a function of the expected scaling $h(E/P)^{1/3}$ ($h$ is the wall thickness, $E$ its elastic modulus and $P$ the inner pressure). Alga cells: *Chara corallina* (square), *Nitella* (triangle pointing down) and *Acetabularia acetabulum* (diamond). Root hair: *Arabidopis thaliana* (triangle pointing up). Yeast: *Saccharomyces cerevisiae* (+ in square). Fungi: *Phycomyces* (\ in square) and *Saprolegnia ferax* (• in circle). Guinea pig cochlear outer hair cell (filled circle). Bacteria: *Magnetospirillum gryphiswaldense* (filled square), *Bacillus subtilis* (filled diamond), *Escherichia coli* (filled triangle pointing up), *Pseudomonas aeruginosa* (filled triangle pointing down) and *Saccharopolyspora erythraea* (• in square). Gas vesicles (included for comparison): *Microcystis spirillum* (+) and *Anabeana flos-aquae* (x). Solid line: best fit to eq. (1) of all data except filled symbols and gas vesicles. **b,** Cell aspect ratio $R/h$ as a function of the modulus/pressure ratio $E/P$. Same symbols as in **a.** Solid lines: best fit of filled symbols to eq. (2) and of all other symbols except gas vesicles to eq.(1).

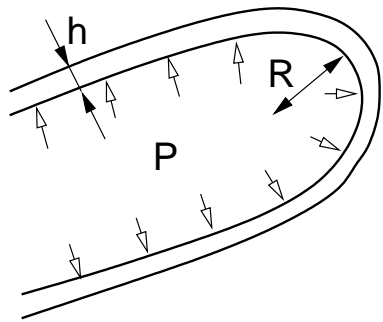

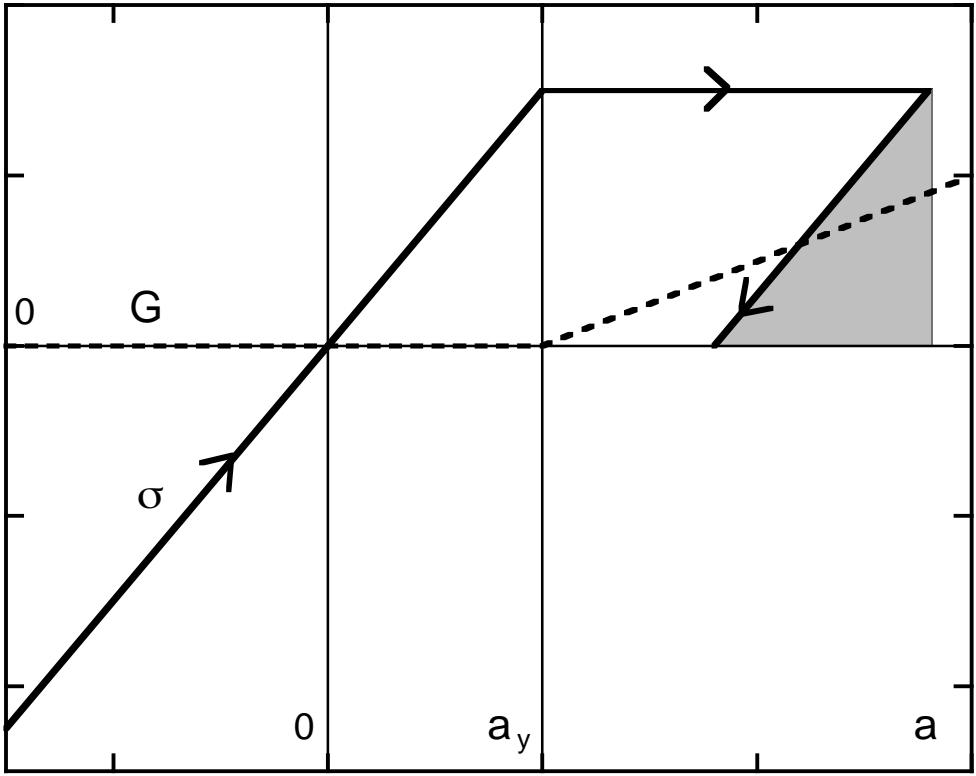

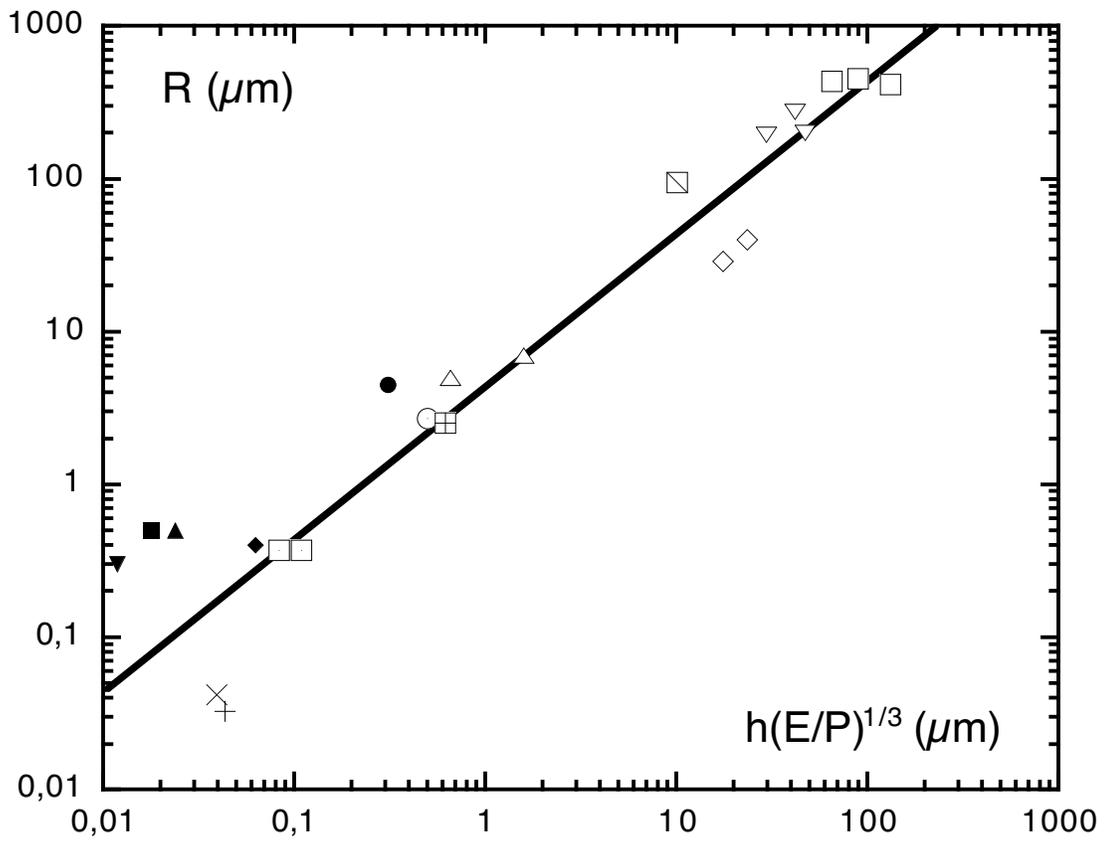

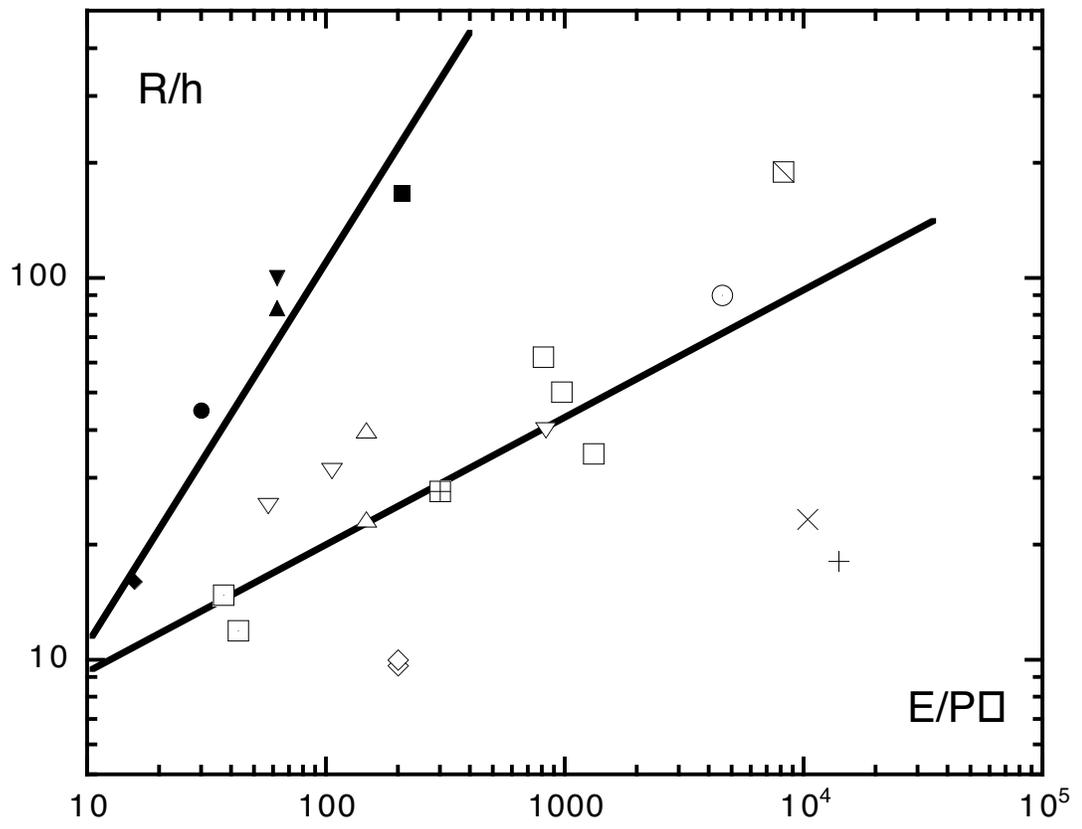